\documentclass[aps,pre,twocolumn,superscriptaddress]{revtex4}

\usepackage[dvips]{graphicx}
\usepackage{amssymb}

\begin{document}

\title{Nonlinear Fano resonance and bistable wave transmission}

\author{Andrey E. Miroshnichenko}

\affiliation{Nonlinear Physics Centre and Centre for Ultra-high
bandwidth Devices for Optical Systems (CUDOS), Research School of
Physical Sciences and Engineering, Australian National University,
Canberra ACT 0200, Australia}

\author{Sergei F. Mingaleev}

\affiliation{Department of Physics, University of Central Florida, 
Orlando, FL 32816, USA}
\affiliation{Bogolyubov Institute for Theoretical Physics, 
Kiev 03143, Ukraine}

\author{Sergej Flach}

\affiliation{Max-Planck-Insitut f\"ur Physik komplexer Systeme,
N\"othnitzerstr. 38, Dresden 01187, Germany}

\author{Yuri S. Kivshar}
\affiliation{Nonlinear Physics Centre and Centre for Ultra-high
bandwidth Devices for Optical Systems (CUDOS), Research School of
Physical Sciences and Engineering, Australian National University,
Canberra ACT 0200, Australia}

\begin{abstract}
We consider a discrete model that describes a linear chain of
particles coupled to a single-site defect with instantaneous Kerr
nonlinearity. We show that this model can be regarded as a
nonlinear generalization of the familiar Fano-Anderson model, and
it can generate the amplitude depended bistable resonant transmission
or reflection. We identify these
effects as the nonlinear Fano resonance, and study its properties for
continuous waves and pulses.
\end{abstract}

\maketitle

\section{Introduction}

The Fano resonance is widely known across many different branches
of physics; it manifests itself as a sharp asymmetric profile of
the transmission or absorption lines, and it is observed in
numerous physical systems, including light absorption by atomic
systems~\cite{uf61}, Aharonov-Bohm interferometer
\cite{junads94,kkhaskyi02} and quantum
dots~\cite{goeres00,bulka01,spinfilters}, resonant light
propagation through photonic-crystal
waveguides~\cite{shfjdj02,shf02,arcjfy03,shfwsjdj03,mfyshfms03,vljpv04},
and phonon scattering by time-periodic scattering
potentials~\cite{etpfb93,swksk01,sfaemvfmvf03}. From the viewpoint
of the fundamental physics, the Fano resonance may appear in the
systems characterized by a certain discrete energy state that
interacts with the continuum spectrum through an {\em
interference effect}. Usually, the discrete state is created by a
defect that allows one (or several) additional propagation paths
in the wave scattering which interact constructively or
destructively. In the transmission line, this interference effect
leads to either {\em perfect transmission} or {\em perfect
reflection},  producing a sharp asymmetric profile.

In the classical paper, Fano~\cite{uf61} derived the general
formula, which describes asymmetric line shape of the transmission
or absorption lines,
\begin{eqnarray}
\label{fano_formula}
\mathcal{F}(\epsilon) = \frac{(\epsilon+f)^2}{\epsilon^2+1}\;,
\end{eqnarray}
where $\epsilon=(E-E_R)/(\gamma_q/2)$ is the dimensionless energy in
the units of the resonance width $\gamma_q$, $f$ is the asymmetry
parameter (Fano factor), and $E_R$ is the resonance energy. In the
limit $f\rightarrow\infty$, this formula can also describe the
so-called Breit-Wigner resonance profile~\cite{junads94}. This
quite universal formula is used usually to fit a particular
profile observed in experiments and, therefore, to provide a proof that the observed
phenomenon can be classified as the Fano resonance.

%
\begin{figure}[htb]
\vspace{20pt} \centerline{
\includegraphics[width=60mm]{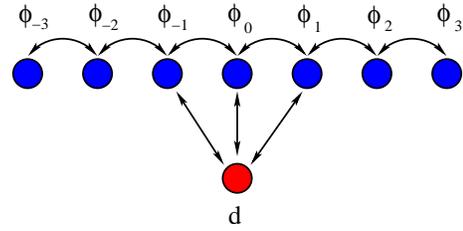}}
\caption{\label{fig1} Schematic of the
Fano-Anderson model. An array of blue circles correspond to a
linear chain and isolated red circle is a defect which can be either
linear or nonlinear. Arrows indicate the coupling between different states.}%
\end{figure}

One of the simplest models that can describe the resonant coupling
and interaction between a discrete state and continuum spectrum is
the so-called Fano-Anderson model~\cite{mahan} (see a sketch in
Fig.~\ref{fig1}). It describes a linear "atomic" chain with 
the nearest-neighbor interaction forces and interacting with
a defect state through the nearest
neighbors. In application to tight-binding models in solids, these
forces are linked to a hopping probability. 
This simple model allows to describe the basic physics
of the Fano resonance in a simple way. In particular, this model
allows some analytical study that may serve as a guideline for the
analysis of more complicated physical models that predict the
Fano resonance effect.

In this paper we consider an important generalization of the
Fano-Anderson model under the assumption that a single-site defect
is nonlinear. We show that this model allows to describe \textit{a
nonlinear Fano resonance}, and it can generate extremely high
contrast between the bistable states in its transmission with low
input power. We study resonant nonlinear transmission and the
properties of the nonlinear Fano resonance for continuous waves and
pulses. We would like to mention that a typical physical situation
when this model can be employed directly for describing the
resonant effects is the light transmission in waveguides and
waveguide junctions created in two-dimensional photonic crystals
with embedded high$-Q$ resonators (defects or cavities) with
nonlinear response. In the nonlinear regime, a defect supports one
(or several) localized state that is characterized by a discrete
eigenvalue. When the excited localized state interacts with the
photonic-crystal waveguide, the system can display interference
effects and Fano resonance.

The aim of this paper is twofold. First, we study both linear and
nonlinear transmission in discrete Fano-Anderson models and obtain a number of
explicit analytical results which allow to get a deeper
understanding of the physics of this phenomenon. Second, we
analyze in detail the effect of nonlinearity on Fano resonance and
demonstrate that, first, it leads to a shift of the Fano resonance
frequency and, second, it can lead to bistability in the wave
transmission.

The paper is organized as follows. In Section II we describe our
discrete nonlinear model which provides a generalization of the
Fano-Anderson model. In Section III, we describe the main features
of the Fano resonance for the case of a local coupling and then
study the effect of nonlinearity of the defect on the
resonant transmission. In particular, we define the conditions for
the bistable transmission and show that the nonlinear Fano resonance
can exist in a broad frequency range. Section IV is devoted to the
study of a model with nonlocal coupling in both linear and
nonlinear regimes. In Section V we consider the wavepacket scattering.
Section VI concludes the paper.

\section{Model}

One of the simplest models which describes the physics and the main
features of the Fano resonance is the so-called Fano-Anderson
model~\cite{mahan}. In this paper, we use a modified
version of these model described by the following
Hamiltonian (see Fig.~\ref{fig1})
\[
H=\sum\limits_nC \phi_n\phi^{*}_{n-1}+ E_d|d|^2+\lambda
|d|^4+d^{*}\sum\limits_jV_j\phi_j+ {\rm c.c.},
\]
where the asterisk denotes the complex conjugation. This model
describes the interaction of two subsystems. One sub-system is a
straight {\em linear} chain with the complex field amplitude
$\phi_n$ at site $n$ which are coupled by nearest-neighbor coupling $C$; this
is the system characterized by the frequency band of the continuum
spectrum. The second subsystem
consists of an additional discrete state $d$ with the local
energy value $E_d$. For the nonlinear model, we assume that the
defect possesses a cubic nonlinear response, $\lambda$ being the
nonlinear parameter. The interaction between these two subsystems is
described by the coupling coefficients $V_j$.

From the lattice Hamiltonian, we derive a system of coupled
nonlinear dynamic equations
\begin{eqnarray}
\label{eq2}
i\dot{\phi}_n&=&C(\phi_{n-1}+\phi_{n+1})+ d \sum\limits_jV_j\delta_{nj}\;,\nonumber\\
i\dot{d}&=&E_dd+\lambda|d|^2d+\sum\limits_jV_j\phi_j,
\end{eqnarray}
where the dot stands for the derivative in time. 
For further analysis, we look for
stationary solutions of this system in the form,
\begin{equation}
\label{eq3} \phi_n(\tau) =A_n\mathrm{e}^{-\mathrm{i}\omega \tau}, \;\;\; d(\tau
)=B\mathrm{e}^{-\mathrm{i}\omega\tau}
\end{equation}
which allow us to describe \textit{elastic} scattering process 
by means of a system of nonlinear algebraic equations,
\begin{eqnarray}
\label{eq4}
\omega A_n&=&C(A_{n-1}+A_{n+1})+ B\sum\limits_jV_j \delta_{nj}\;,\nonumber\\
\omega B&=&E_dB+\lambda|B|^2B+\sum\limits_jV_j A_j\;.
\end{eqnarray}

The model (\ref{eq4}) is the main subject of our analysis of the
stationary wave scattering. In the following two sections, we
study first the regime of a local coupling when only one coupling
coefficient $V_0$ is nonzero, and after that consider a nonlocal
coupling with up to three nonzero coupling coefficients $V_{-1,0,1}$. In both
these cases, we start our analysis from the linear regime and then
analyze the Fano scattering in the nonlinear regime when $\lambda
\neq 0$. A full time-dependent model (\ref{eq2}) is discussed in
Section V of the paper when we consider the wavepacket
scattering.

\section{Local coupling}

First, we analyze the case of a local coupling when $V_j\equiv0$
for $j\neq0$ and $V_0$ is nonzero and study separately the linear ($\lambda
=0$) and nonlinear ($\lambda \neq 0$) regimes.

\subsection{Linear scattering}

We consider the scattering of plane waves with the dispersion
$\omega_q=2C\cos q$ propagating along the linear chain. Using the
second equations of the system (\ref{eq4}), we find a simple link
between two defect-site characteristics
\begin{eqnarray}
\label{discrete_state}
B=\frac{V_0 A_0}{\omega-E_d}
\end{eqnarray}
and obtain a single equation
\begin{eqnarray}
\label{eq5}
\omega A_n=C(A_{n-1}+A_{n+1})+\frac{V_0^2
A_0}{\omega-E_d}\delta_{n0}\;,
\end{eqnarray}
with a one-site scattering potential. For the scattering problem,
we consider the following boundary conditions
\begin{eqnarray}
\label{boundary}
A_n = \left\lbrace %
 \begin{array}{lc}
    I\mathrm{e}^{\mathrm{i}qn}+r\mathrm{e}^{-\mathrm{i}qn},& n<0,\\
    t\mathrm{e}^{\mathrm{i}qn}, & n> 0,
   \end{array}
   \right.
\end{eqnarray}
where $I$, $r$, and $t$ have the meaning of the incoming, reflected
and transmitted wave amplitudes far from the defect site.
Without loss of generality, we will assume that incoming amplitude $I$
is real.
According to Eq.~(\ref{eq5}), strength of the scattering
potential depends on the incoming frequency $\omega$, and the
system should demonstrate {\em resonant scattering}. If the
frequency of the defect is placed in the propagation frequency
band of the chain, i.e. $|E_d| \le2 C$,  the scattering potential
in Eq.~(\ref{eq5}) becomes infinitely large at $\omega\equiv\omega_F=E_d$, and
this will lead to total reflection of the incoming wave.

%
\begin{figure}[htb]
\vspace{20pt} \centerline{
\includegraphics[width=80mm]{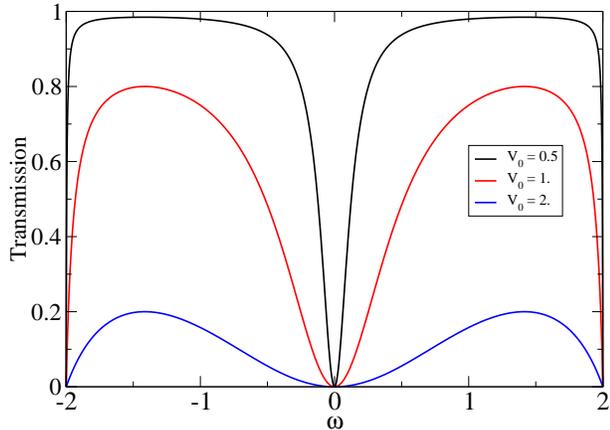}
} \caption{\label{fig2}Transmission coefficient (\ref{eq7})
calculated for the linear Fano-Anderson model (\ref{eq2}) in the
case $E_d=0$,
$C=1$, and $\lambda=0$, for several values of the coupling coefficient $V_0$.}%
\end{figure}

Using the well-known transfer-matrix approach, we obtain the
analytical result for the transmission coefficient defined as
$T=|t|^2/I^2$
\begin{eqnarray}
\label{eq7}
T=\frac{\alpha_q^2}{\alpha_q^2+1}\;,
\end{eqnarray}
where $\alpha_q=c_q(\omega_q-E_d)/V_0^2$ and $c_q=2C\sin q$. 
This result
corresponds to a simple physics: For the resonant frequency
$\omega_F$, there exist two scattering channels, with and without
the defect state excited, and {\em destructive interference}
between the waves passing these channels leads to a complete
suppression of the wave transmission (see Fig.~\ref{fig2}). As a
matter of fact, the result (\ref{eq7}) corresponds to the Fano
formula (\ref{fano_formula}), where $\alpha_q$ plays a role of the
dimensionless energy, $E_d$ is the resonant energy eigenvalue, and
the resonance width is defined as
\begin{eqnarray}\label{width}
\Gamma=\frac{V_0^2}{C\sin q_F}\;,
\end{eqnarray}
where $q_F$ is a wavenumber at the resonance, $\omega_F =
\omega(q_F)$.  According to this result, the width is proportional
to the coupling strength $V$, and it depends on the position of
the Fano resonance with respect to the boundary of the
spectrum~\cite{aemsfbm03}. In this case, the asymmetry parameter
$f$ vanishes, and the transmission profile is symmetric.

Transmission vanishes when $\alpha_q=0$; this happens at the edges
of the continuous spectrum band, $q=0$ and $q=\pi$, due to
the vanishing of group velocities, and at the
resonant frequency $\omega_F$ which, in the case of a local
coupling, coincides with the discrete eigenvalue of the defect
mode, $\omega_{F}=E_d$. If the defect is coupled to two or more
sites, the resonant frequency is renormalized, as discussed below.
We notice that when the resonant frequency is in the middle of the
propagation band, the transmission profile is symmetric (see
Fig.~\ref{fig2}), and in the case of a local coupling the Fano
resonance is observed as the resonance reflection.

%
\begin{figure}[htb]
\vspace{20pt} \centerline{
\includegraphics[width=80mm]{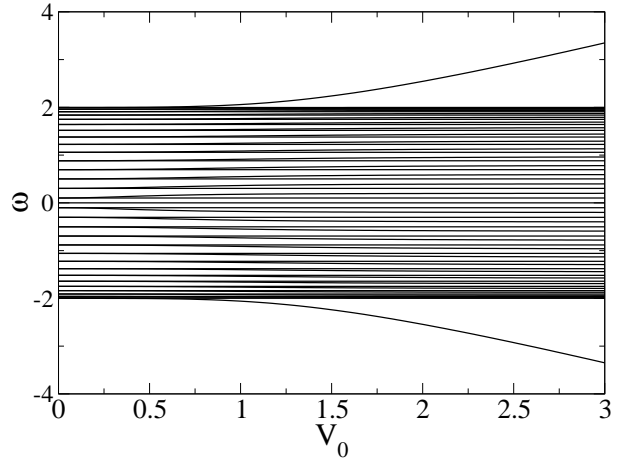}
} \caption{\label{fig3} Frequency spectrum of the system
(\ref{eq4}) vs. the coupling parameter $V_0$ for the case $C=1$,
$E_d=0$, and $\lambda=0$.}
\end{figure}

Using that $|A_0|^2=TI^2$ and the relation
(\ref{discrete_state}), we can calculate analytically the
amplitude of the defect state, that riches the maximum
\begin{eqnarray}\label{q-fac}
|B_{\rm max}|^2= I^2/2\Gamma
\end{eqnarray}
at the Fano resonance frequency $\omega_F$.

When the coupling coefficient between the defect and the chain
vanishes, $V_0=0$, linear modes of the system (\ref{eq4}) are
described by the continuous spectrum $\omega_q=2C\cos q$, while
the defect with the frequency $\omega =E_d$ is uncoupled. For a
finite coupling $V_0\not=0$, the defect generates two local modes
(or bound states) with the frequencies bifurcating from the upper
and lower edges of the continuous spectrum, $|\omega_L|\ge 2C$, as
shown in Fig.~\ref{fig3}. 
The appearance of these local modes does not show any characteristic
feature in transmission close to the band edges.
For the case $E_d=0$, these states are
located symmetrically, because the defect frequency is at the
middle of the spectrum $\omega_q$.

In general, the frequency of the localized state $\omega_L$ 
of the one-site potential is
given by the expression $\omega_L^2=4C^2+\xi^2$, where $\xi$ is
the strength of the potential, and in our case,
\begin{eqnarray}
\label{eq6} \omega_L^2 = 4C^2+\frac{V_0^4}{(\omega_L-E_d)^2}.
\end{eqnarray}
Formally, this is the fourth-order equation in $\omega_L$, but for
$E_d$ inside the spectrum $\omega_q$, there exist only two real
solutions which correspond to two bound states.
%
%
%
\begin{figure}[htb]
\vspace{20pt} \centerline{
\includegraphics[width=80mm]{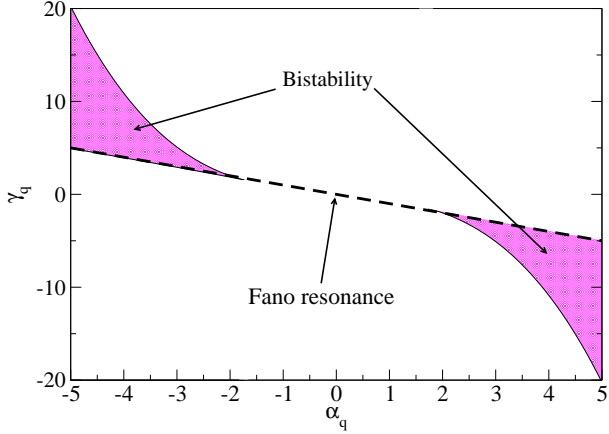}}
\caption{\label{fig4}Parameter plane ($\alpha_q$,$\gamma_q$) defined
by different solutions of Eq.~(\ref{cubic}). Filled areas
correspond to bistability, a straight dashed line at the boundary
of the bistability region corresponds to the Fano resonance.}
\end{figure}

\subsection{Nonlinear scattering and bistability}

Now we assume that the isolated defect is nonlinear, i.e. $\lambda
\neq 0$. Using Eqs. (\ref{discrete_state}) and the continuity condition
at the defect site, $I+r=t$, we obtain the general expression for
the transmission coefficient,

\begin{eqnarray}
\label{eq12}
T=\frac{x^2}{x^2+1}\;,
\end{eqnarray}
where $x$ is a real solution of the cubic equation

\begin{eqnarray}
\label{cubic}
(x^2+1)(x-\alpha_q)-\gamma_q =0,
\end{eqnarray}
with the parameter
$\gamma_q=\lambda c_q^3I^2/V_0^4$.

The transmission coefficient defined by Eq.~(\ref{eq12})
corresponds again to the general Fano formula
(\ref{fano_formula}). During our derivation, we put
$x=\tan \theta$, where $\theta = \mathrm{Arg}(r)$ 
is a phase of the reflection amplitude. Due to the local
range of the scattering potential (\ref{eq5}), 
it can be shown  \cite{lipkin} that the
phase of the reflection amplitude is equal to 
$\theta=\delta(q)+\pi/2$,
where $\delta(q)$ is a scattering phase shift.
Therefore, we can consider eq.(\ref{cubic})
as a nonlinear equation for the scattering phase shift,
with $x=\cot\delta(q)$. Note here, that eq.(\ref{cubic})
may support complex solutions for scattering phase shift, which
will correspond to \textit{inelastic} scattering \cite{lipkin}.
Such situation is beyond the scope of this paper.
The expression (\ref{eq12}) is valid only for real solutions
of eq.(\ref{cubic}), which correspond to elastic scattering process.
 
Transmission (\ref{eq12}) vanishes at $x=0$ when
$\alpha_q=-\gamma_q$ :
\begin{eqnarray}
\label{zero}
V_0^2(E_d-\omega_q)+\lambda I^2(4C^2-\omega_q^2)=0
\end{eqnarray}
or $c_q=0$.

%
%
\begin{figure}[htb]
\vspace{20pt}
\centerline{
\includegraphics[width=80mm]{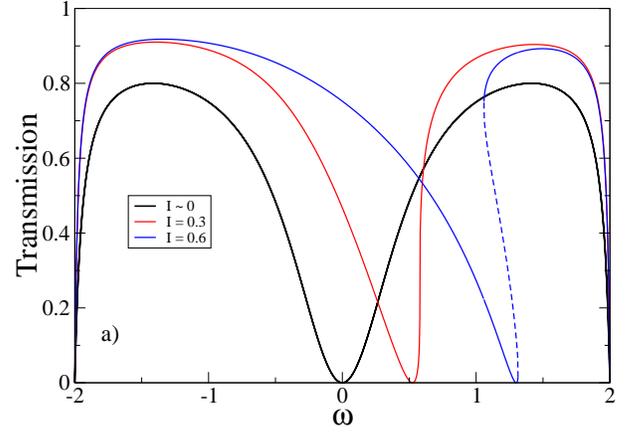}}
\vspace{30pt} \centerline{
\includegraphics[width=80mm]{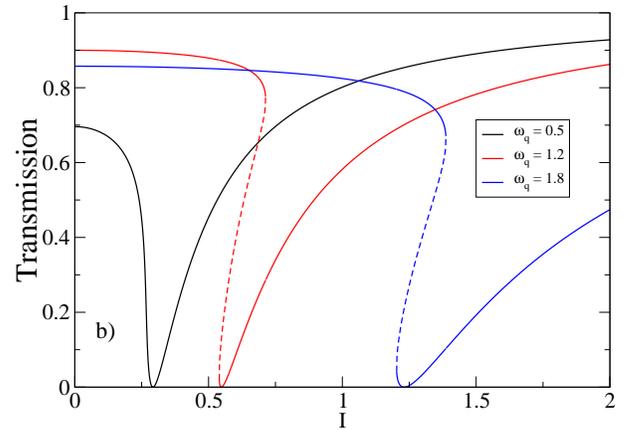}}
\caption{\label{fig5} Dependence of the transmission coefficient
on (a) the wave frequency $\omega_q$ for a fixed intensity $I$,
and (b) input intensity $I$ for a fixed frequency $\omega$, for
$C=\lambda=1$, $V_0=0.8$ and $E_d=0$. Bistability regions
($|\alpha_q|>\sqrt{3}$) are denoted by dashed lines.}
\end{figure}

When $\gamma_q=0$, there exists only one real solution  of
Eq.~(\ref{cubic}), $x=\alpha_q$, it leads to the result (\ref{eq7})
obtained above. In the nonlinear regime ($\gamma_q\neq0$), there
exist up to three real solutions, the output may become a
multi-valued function of the input, and this may lead to {\em
bistability} of the wave transmission \cite{gibbs}.

To define the regions where such a bistability transmission can
occur, we present the left-hand side of Eq.~(\ref{cubic}) in the
form $F(x)=f(x)-\gamma_q$, so that the bistability regions are
determined by local extrema of the function $f(x)$. When
$|\alpha_q|<\sqrt{3}$ there is no local extrema, and for any value
of $\gamma_q$ (and also $\lambda$ or $I$) there exists only one real
solution of Eq.~(\ref{cubic}) and therefore the transmission is
always single-valued. For $|\alpha_q|>\sqrt{3}$, a pair of minimum
and maximum points of the function $f(x)$ appear at $x_{\rm min,
max}=(\alpha_q\pm\sqrt{\alpha_q^2-3})/3$, respectively, and the
bistability region is determined by the condition,

\begin{eqnarray}
\label{bistability} f(x_{\rm min}) < \gamma_q < f(x_{\rm max}),
\end{eqnarray}
filled on the parameter plane $(\alpha_q,\gamma_q)$ in
Fig.~\ref{fig4}.

In the nonlinear regime, the transmission coefficient depends on
two parameters: the wave frequency $\omega_q$ and input intensity
$I$. First, we study the dependence of the transmission
coefficient on the wave frequency. From the second equation of
Eqs.~(\ref{eq4}), we can see that nonlinearity shifts the
frequency of the localized mode and, therefore, it shifts the
position of the Fano resonance. By denoting the left-hand side
of the Eq.~(\ref{zero}) as $g(\omega_q)$, it follows
that $g(\omega_q)$ is a quadratic function of $\omega_q$.
If we put $E_d=0$ then $g(-2C)g(2C)<0$ and, therefore, 
there is an unique solution of $g(\omega_q)$ inside the interval $[-2C,2C]$. 
In the limit of large intensity, i.e. for $I
\rightarrow \infty$, the solutions of the equation $g(\omega_q)=0$
approach the values $\pm 2C$. Thus, for {\em any value} of the
input intensity $I$ there always exists a single Fano resonance
[see Fig.~\ref{fig5}(a)].

To define the position of the Fano resonance analytically, we
solve  Eq.~(\ref{zero}) explicitly with respect to the frequency
$\omega_q$,
\begin{eqnarray}
\label{nlF}
\omega_F=-\frac{V_0^2\pm\sqrt{V_0^4+4\lambda I^2(V_0^2E_d+4\lambda I^2C^2)}}{2\lambda I^2}\;.
\end{eqnarray}
A proper sign in Eq.~(\ref{nlF}) should be chosen depending on the
particular values of the parameters.

Thus, by increasing the input intensity, we shift the position of
Fano resonance such that for $I >I_{\rm cr}$, the transmission
coefficient becomes multi-valued and a bistability region appears.
Because both $\alpha_q$ and $\gamma_q$ depend on the frequency
$\omega_q$, a change of the frequency corresponds to a move in the
parameter space $(\alpha_q, \gamma_q)$. For
$E_d=0$,  $\alpha_q$ changes in the interval $[-\alpha_q^{\rm
max},\alpha_q^{\rm max}]$, where
\begin{eqnarray}\label{alpha_max}
\alpha_q^{\rm max}= 2C^2/V_0^2
\end{eqnarray}
and bistability is expected for $\alpha_q^{max}>\sqrt{3}$ [see
Fig.~\ref{fig5}(a)].

To study the dependence of the transmission coefficient on the
input intensity $I$, we solve Eq.~(\ref{zero}) and obtain
\begin{eqnarray}
\label{I}
I^2=\frac{(\omega_q-E_d)V_0^2}{\lambda(4C^2-\omega_q^2)}\;.
\end{eqnarray}
Equation (\ref{I}) indicates that, depending on the sign of
nonlinearity $\lambda$, the Fano resonance appears for any
frequency from the interval $[-2C,E_d]$ or $[E_d,2C]$, and the
frequency should satisfy the condition,
\begin{eqnarray}
\label{interval} (E_d-\omega_q)\lambda<0.
\end{eqnarray}

On the plane of parameters $(\alpha_q, \gamma_q)$, only $\gamma_q$
depends on the input intensity $I$ , so that varying $I$ we move
along the line $\alpha_q=const$ and $\gamma_q<0$ or $\gamma_q>0$
(depending on the sign of the nonlinearity parameter $\lambda$).
In order to achieve bistability, we should satisfy the conditions
$|\alpha_q|>\sqrt{3}$ and $\alpha_q \gamma_q<0$. The latter condition
can be reduced to the inequality $(E_d-\omega_q)\lambda<0$, which
defines the existence of the Fano resonance (\ref{interval}). In
other words, bistability appears only simultaneously with the
presence of Fano resonance, see Fig.~\ref{fig5}(b) and
Fig.~\ref{fig4}.

\section{Nonlocal coupling}

As a matter of fact, the local coupling discussed above provides a
relatively simple approach for the derivation of analytical
formulas and understanding the major physics of the Fano
resonance. However, very often the physical problems where the
Fano resonance is observed are more complicated. In particular,
one of the major property of realistic physical models such as
photonic-crystal waveguides coupled to localized defect modes, is
long-range interaction and nonlocal coupling~\cite{mingaleev}.
Below, we study the generalized discrete (linear and nonlinear)
Fano-Anderson model where the defect is coupled to three nearest
neighbors of the chain, i.e. in Eq.~(\ref{eq2}) $V_j \neq 0$  for
$j=-1,0,1$.

\subsection{Linear scattering}

In order to find the transmission coefficient for a nonlocal
model, we modify the boundary conditions,
(\ref{boundary})
\begin{eqnarray}\label{boundary2}
A_n = \left\lbrace %
 \begin{array}{lc}
    I\mathrm{e}^{\mathrm{i}qn}+r\mathrm{e}^{-\mathrm{i}qn}\;,&n<-1,\\
    t\mathrm{e}^{\mathrm{i}qn}\;,&n> 1.
   \end{array}
   \right.
\end{eqnarray}
In addition, we should solve a set of coupled equations for the
sites $n=-1,0,1$ with the boundary conditions (\ref{boundary2})
with respect to the reflection coefficient $R=|r|^2/I^2$,
\begin{eqnarray}
\label{refl_nonloc}
R=\frac{\mathrm{Re}(b_q)^2}{\mathrm{Re}(b_q)^2+[c_q(\omega_q-E_d)+\mathrm{Im}(b_q)]^2}\;,
\end{eqnarray}
where
$b_q=V_{-1}^2+V_0^2+V_1^2+2\mathrm{e}^{2\mathrm{i}q}V_{-1}V_1+2\mathrm{e}^{\mathrm{i}q}V_0(V_{-1}+V_1)$.
The transmission coefficient $T$ can be found
from the identity $T+R=1$.

In the nonlocal case, the transmission coefficient displays more
complex behavior as compared with the case of a one-side coupling
(see Sec.~III above). In particular, in this case the Fano
resonance is asymmetric. Perfect reflection occurs when the second
term in the denominator of Eq.~(\ref{refl_nonloc}) vanishes. The
corresponding condition can be rewritten in the form
\[
c_q[C(\omega_q-E_d)+V_0(V_{-1}+V_1)+2\cos qV_1V_{-1}]=0
\]
and it is satisfied at the spectrum band edges $q=0$ and $q=\pi$,
due to zero group velocities
and also when
%
\begin{figure}[htb]
\vspace{20pt} \centerline{
\includegraphics[width=80mm]{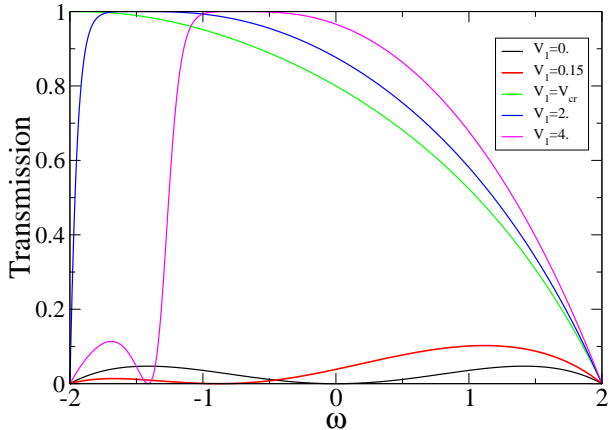}
} \caption{\label{fig7}Transmission coefficient $T$ vs. frequency
$\omega_q$ for the three-site interaction, for different values of
$V_1=V_{-1}$. Other parameters are: $C=1$, $V_0=3$ and $E_d=0$. According
to Eq.~(\ref{perf_nonloc3_edge}), the perfect transmission occurs
inside the spectrum at $V_1=1.5$. There is an interval for
coupling coefficient $V_1$, when there is no perfect reflection
inside the spectrum. For larger values of $V_1$, both perfect
transmission and perfect reflection move to the middle of the
spectrum $\omega_q=0$.
}%
\end{figure}

%
\begin{eqnarray}
\label{fano_nonloc}
\omega |_{T=0} = \frac{C^2E_d-C V_0(V_{-1}+V_1)}{C^2+V_1V_{-1}}\;.
\end{eqnarray}

Equation~(\ref{fano_nonloc}) shows that, in the case of nonlocal
coupling to a defect state, the position of the perfect reflection
gets shifted. If the coupling is strong enough, the resonance can
move outside the spectrum band.

In addition, there exists the frequency when a perfect
transmission occurs. This happens when $\mathrm{Re}(b_q)=0$ or
\begin{eqnarray}\label{perf_nonloc}
\frac{\omega }{C}|_{T=1}=\frac{-V_0(V_1+V_{-1})\pm
\sqrt{(V_0^2-4V_1V_{-1})(V_1-V_{-1})^2}}{2V_1V_{-1}}\;
\end{eqnarray}
and the corresponding frequency $\omega|_{T=1}$ does not depend
on the discrete state energy $E_d$. This means when the position
of perfect reflection $\omega|_{T=0}$ is shifted, the point of
the perfect transmission remains unchanged. This property allows
us to vary the width of the asymmetric Fano resonance by changing
the energy $E_d$.

Formally, there exist two solutions for the perfect transmission.
But in real systems, such as waveguides in photonic crystals, some
symmetries of the coupling coefficients hold, i.e. $V_1=V_{-1}$ or
$V_{-1}=V_0$ and $V_1=0$, in the cases of three and two nonzero
coupling terms, correspondingly. For the first case $V_1=V_{-1}$,
we obtain
\begin{eqnarray}
\label{perf_nonloc3} \omega |_{T=1}=-C (V_0/V_1)
\end{eqnarray}
and therefore only one solution for the perfect transmission is
possible. For small values of $V_1$, there is no perfect
transmission, and it takes place at the boundary of the spectrum
$\omega |_{T=1}=-2C$, when
%
%
\begin{figure}[htb]
\vspace{20pt}
\centerline{
\includegraphics[width=80mm]{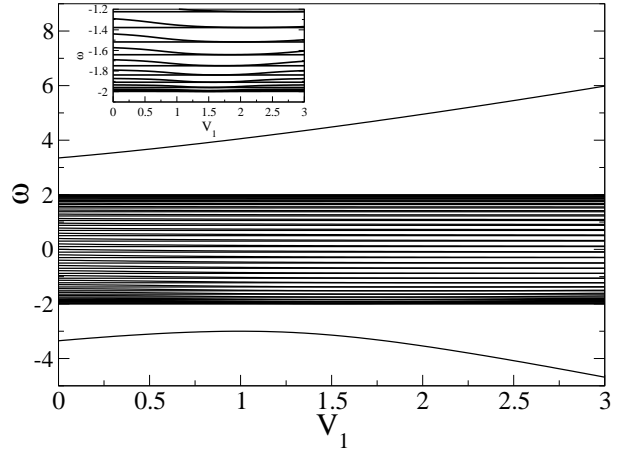}
} \caption{\label{fig8}Frequency spectrum for the nonlocal model
vs. coupling coefficient $V_1$ for the same values of parameters
as in Fig.~\ref{fig7}. The insert shows a zoomed area with the
double degeneracy of the energy levels near the critical value
$V_{\rm cr}$.
}%
\end{figure}

\begin{eqnarray}
\label{perf_nonloc3_edge}
V_1=V_0/2
\end{eqnarray}
and it exists for any larger value of $V_1$. When
$V_1\rightarrow\infty$, the frequency of the perfect transmission
moves to the middle of the spectrum, $\omega |_{T=1}\rightarrow0$
(see Fig.~\ref{fig7}).

In the nonlocal model, the linear spectrum varies substantially as
a function of the coupling coefficient $V_1$, see Fig.~\ref{fig8}.
For small values of $V_1$, two localized states exist outside the
linear spectrum band $\omega_q$ (cf. Fig.~\ref{fig3}). The
frequency of the upper localized state goes away from the spectrum
for all $V_1$. The lower localized state approaches the spectrum
band but then deviates again for larger $V_1$.

Levinson's theorem \cite{bnzaas90} allows us to connect some properties
of the linear spectrum (Fig.\ref{fig8}) with transmission coefficient (Fig.\ref{fig7}).
According to this theorem scattering phase shift at one of the band edge
can be represented as
\begin{eqnarray}\label{phaseshift}
\delta(q=0)=\pi(N_b+l/2)\;,
\end{eqnarray} 
where $N_b$ is a number of bound state of the system and $l$
is a number of quasi-bound states at the band edge.
Using the relation between transmission coefficient and 
scattering phase shift\cite{lipkin,bnzaas90},
it can be shown that $T(q=0)=0$ for even and $T(q=0)=1$ 
for odd values of $l$.
Therefore, we can expect
that some quasi-bound states should appear at the band edge, when
the condition (\ref{perf_nonloc3_edge}) is satisfied. The insert
in Fig.~\ref{fig8} shows that exactly for this value of the
coupling parameter $V_1$, the spectrum band edge becomes doubly
degenerate. In this situation $l=1$.

%
\begin{figure}[htb]
\vspace{20pt} \centerline{
\includegraphics[width=80mm]{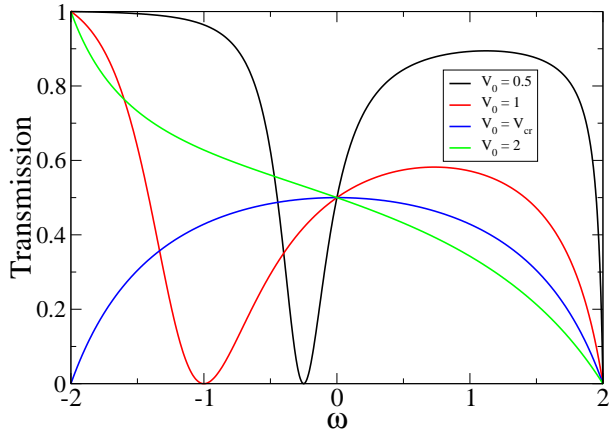}
} \caption{\label{fig9}Transmission coefficient $T$ vs. frequency
$\omega_q$ for the nonlocal model with two nonzero couplings for
different values of coupling $V_0=V_{-1}$. Other parameters are $C=1$ and
$E_d=0$. Perfect transmission occurs at $\omega_1=-2$. When
$V_0=V_{\rm cr}=\sqrt{2}$, the perfect reflection state split off
the spectrum band and transmission at $\omega=-2$ vanishes.
}%
\end{figure}

For the second case $V_0=V_{-1}$ and $V_1=0$, we obtain
the following result
\begin{eqnarray}
\label{perf_nonloc2} \omega |_{T=1}=-2C,
\end{eqnarray}
so that the position of perfect transmission does not depend on
the coupling coefficients and it should take place at the boundary
of the spectrum band. According to Levinson's theorem, this may
correspond only when some quasi-bound state exists {\em exactly}
at the band edge, $l=1$ in (\ref{phaseshift}).

\begin{figure}[htb]
\vspace{20pt} \centerline{
\includegraphics[width=80mm]{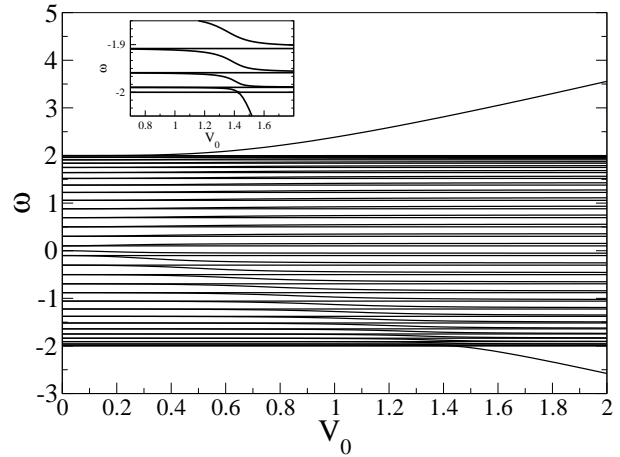}
} \caption{\label{fig10}Frequency spectrum for the nonlocal model
with two nonzero coupling terms $V_0=V_{-1}$ vs. the
coupling coefficient $V$ for the same parameters as in
Fig.~\ref{fig9}. Insert shows a localized state at $\omega=-2$,
which does not interact with other modes. The second bound state
appears at $V_0=V_{\rm cr}$ when the perfect reflection splits off
the spectrum band.
}%
\end{figure}

Figure~\ref{fig9} predicts the perfect transmission at
$\omega_q=-2C$ for all values of $V_0$, except the point $V_0=V_{\rm
cr}$ when the perfect reflection (\ref{fano_nonloc}) splits off
the spectrum band (see Fig.~\ref{fig9}). In this case, for any
nonzero value of $V_0$ there exists an upper localized state and the
quasi-bound state at the band edge $\omega=-2C$ (see Fig.~\ref{fig10}).
In addition to this, at $V_0=V_{\rm cr}$  another quasi-bound state
appears at the band edge of the spectrum. 
In such a situation there are two quasi-bound states ($l=2$)
and according to Levinson's theorem (\ref{phaseshift})
transmission coefficient vanishes at the bottom band edge. 
For larger values of coupling $V_0$ a second bound state exists in the system.

\subsection{Nonlinear scattering}
%

In the nonlinear regime, the reflection coefficient can be found
in the following analytical form

\begin{eqnarray}
\label{fano_nonloc_nl} R = \frac{[{\rm Re}(a_q^2)+ {\rm
Im}(a_q^2)x]^2}{[{\rm Re}(b_q)]^2(1+x^2)}\;,
\end{eqnarray}
where $x$ is a solution of the cubic equation,
%
\begin{eqnarray}
\label{cubic2} \lambda I^2c_q^3y_x^3-(x^2+1){\rm Re}(b_q)^2(|a_q|^2{\rm Re}(b_q)z_x+e_qy_x)=0,\;\;\;
\end{eqnarray} 
where
$y_x = {\rm Re}(a_q^2)+ {\rm Im}(a_q^2)x$,
$z_x = {\rm Re}(a_q^2)x-{\rm Im}(a_q^2)$,
$a_q = V_{-1}+e^{iq}(V_0+e^{iq}V_1)$,
$e_q=|a_q|^2[(\omega_q-E_d)c_q+ {\rm Im}(b_q)]$,
and the symbols ${\rm Re}$ and ${\rm Im}$ stand
for the real and imaginary parts, respectively.

We would like to mention again that the general result for the
reflection coefficient (\ref{fano_nonloc_nl}) is similar to the
Fano formula (\ref{fano_formula}), where $x$ can be understood as
the effective dimensionless energy $\epsilon$ and the ratio ${\rm
Re}(a_q^2)/{\rm Im}(a_q^2)$ is related to the asymmetry parameter
$f$. This formula (\ref{fano_nonloc_nl}) includes all other cases
discussed above. The condition for the perfect transmission ${\rm
Re}(b)=0$ coincides with the result (\ref{perf_nonloc}), so that
the frequency of the perfect transmission does not depend on the
input intensity $I$, nonlinearity parameter $\lambda$, and the
energy of the discrete state $E_d$. The position of the perfect
reflection does depend on all these parameters, and this gives us
with a possibility to vary the width of the asymmetric resonance
in the nonlocal nonlinear regime.

\begin{figure}[htb]
\vspace{20pt}
\centerline{
\includegraphics[width=80mm]{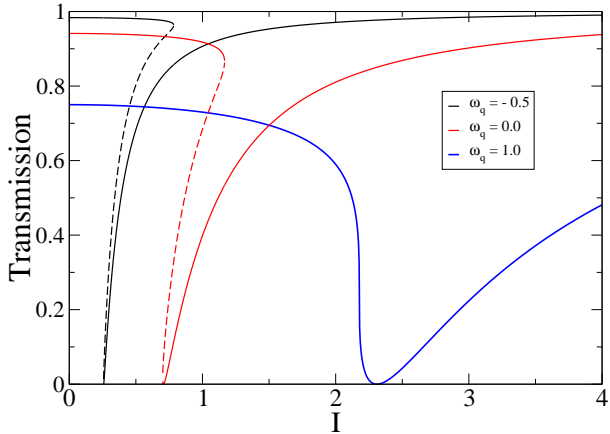}
} \caption{\label{fig11}Transmission coefficient vs. intensity $I$
for the nonlocal nonlinear model for $V_{-1}=V_0=V_1=1$,
$\lambda=1$, $C=1$, and $E_d=0$. Perfect reflection occurs at the
edge of the bistability domain (see Fig.~\ref{fig4}).
Dashed lines correspond to bistable regions.
}%
\end{figure}

Finally, in Fig.~\ref{fig11} we plot the dependence of the
transmission coefficient vs. the input intensity $I$ for different
values of the frequency. These results are similar to those
obtained for the model with a local coupling, presented in
Fig.~\ref{fig5}(b).

\section{Wavepacket scattering}

The previous sections are devoted to the problem of the
time-independent plane wave scattering. In order to verify how
these results can be used for describing realistic wave
scattering, we should analyze the time-dependent scattering of
wavepackets. In this section, we study this problem numerically in
the framework of the time-dependent model (\ref{eq2}).
%

We consider the propagation of a Gaussian wavepacket through a
discrete chain with the defect described by the system of
equations (\ref{eq2}) in the regime of a local coupling, when
$V_0=V$ and $V_{n \neq 0}=0$. As the input wave, we take a
Gaussian wavepacket of the following form
\begin{eqnarray}
\label{gaussian} \phi_n(0) = I_w
\exp\left[-\frac{(n-n_0)^2}{\sigma^2}\right]\exp[-iq_w(n-n_0)],
\end{eqnarray}
where $q_w$ is the carrier wavenumber of the wavepacket, $I_w$ is
its maximum amplitude, $\sigma$ is the spatial width, and $n_0$ is
the initial position. The wavenumber $q_w$ determines the velocity
of the wavepacket.

For  calculating the transmission and reflection coefficients, we
use the conservation of the norm, and define the coefficients as follows
\begin{eqnarray}
\label{scat_amp}
T=\frac{\sum\limits_{n>0}|\phi_n(\tau^*)|^2}{\sum\limits_n|\phi_n(0)|^2}\;,\;\;
R=\frac{\sum\limits_{n<0}|\phi_n(\tau^*)|^2}{\sum\limits_n|\phi_n(0)|^2}\;,
\end{eqnarray}
where $\tau^*$ corresponds to some time after interaction of the
wavepacket with a defect.

\begin{figure}[htb]
\vspace{20pt} \centerline{
\includegraphics[width=80mm]{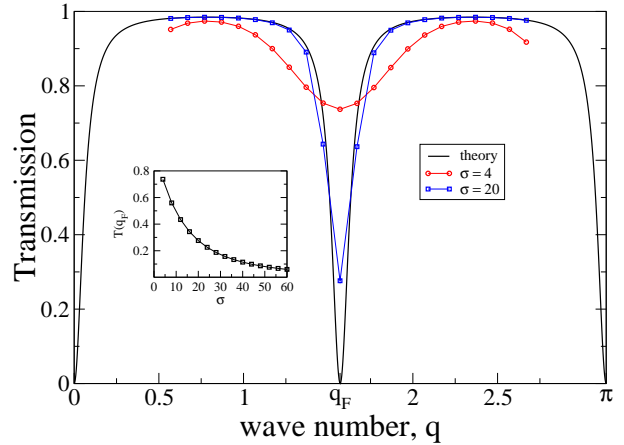}
} \caption{\label{fig14}Transmission coefficient
of the wavepacket in linear regime $\lambda=0$ for different
values of the width $\sigma$. Other parameters are $C=1$, $V=0.5$, $E_d=0$.
For given parameters the spectral width of 
the wavepacket becomes smaller the width of
the resonance $\Gamma=1/4$ for $\sigma>20$.
It leads to a good agreement with the analytical result.
(Insert) Transmission at the
position of the Fano resonance $q_F$ decays exponentially with the spatial width $\sigma$.
}%
\end{figure}

In our simulations, we take $E_d=0$ and
therefore $q_F=\pi/2$. First, we compare numerical results for the wavepacket scattering
with plane-wave analysis in the linear regime (see Fig.\ref{fig14}).
When the spectral width of the wavepacket $2\pi/\sigma$ is
smaller then the width of the resonance (\ref{width}), we have observed 
a good agreement. 
%

\begin{figure}[htb]
\vspace{10pt} 
\centerline{
\includegraphics[width=80mm]{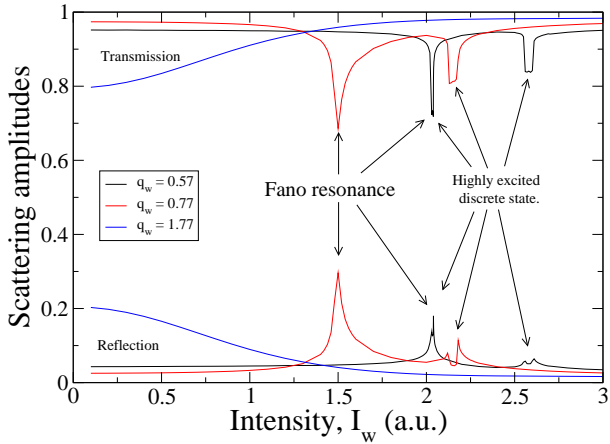}
} \caption{\label{fig12}Transmission and reflection coefficients
vs. the input amplitude $I_w$ of the wavepacket defined in
Eq.~(\ref{gaussian}) for $C=1$, $V=0.5$, $E_d=0$, $\lambda=1$,
$\sigma=4$, and $n_0=-10$.
}%
\end{figure}

In the nonlinear regime, our numerical results 
for the wavepacket scattering are summarized in
Fig.~\ref{fig12} for the dependence of the transmission and
reflection coefficients on the input wavepacket amplitude $I_w$
defined in Eq.~(\ref{gaussian}) for different values of the
carrier wavenumber.  For the case $q_w<q_F$, Fig.~\ref{fig12}
indicates the existence of the Fano resonance characterized by the
resonant deep in the transmission coefficient and simultaneous
resonant peak in the reflection coefficient. When $q_w>q_F$, no
Fano resonance exists. These results provide a good qualitative
agreement with the plane-wave analysis presented above.

\begin{figure}[htb]
\vspace{20pt} \centerline{
\includegraphics[width=80mm]{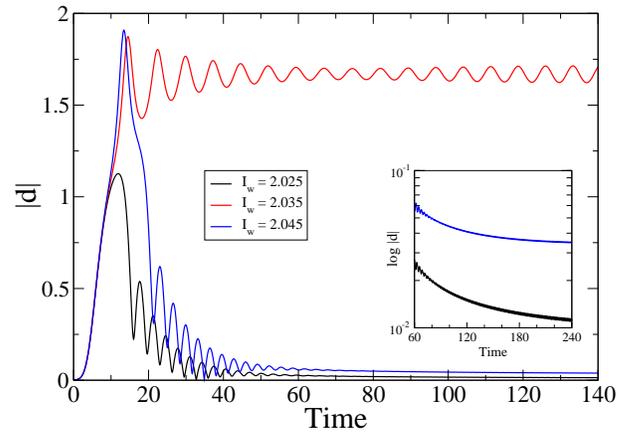}
} \caption{\label{fig13} Evolution of the defect site during and
after the wavepacket scattering, for $q_w=0.57$. Other parameters
are the same as in Fig.~\ref{fig12}.  Different final states of
the defect are observed for a slight variation of the amplitude of
the incoming wavepacket, being a signature of the system
bistability. Inset shows a log-plot of this dependence and
indicates that the defect state relaxes to the ground state
algebraically.
}%
\end{figure}

In addition to this feature associated with the Fano resonance,
there are observed some regions in Fig.~\ref{fig12} when {\em
both} transmission and reflection coefficient decrease, i.e. their
sum does not equal to unity. Usually, the discrete localized
mode becomes excited during the interaction of a wavepacket with
the defect, and then it relaxes to the ground state. For these
particular values of the wavepacket amplitude $I_w$, both defect
site and zero-site particle of the chain get highly excited with long-lived
oscillations after the interaction (see Fig.~\ref{fig13}).
Such a behavior is very similar to the scattering by a delta-potential
with an excited state \cite{lipkin} and it corresponds to the \textit{inelastic}
scattering process. 
In our situation the localized state is excited due to the nonlinear
interaction.

For the
wavenumber $q_w=0.57$ such a situation even happens near the Fano
resonance, i.e. by varying the amplitude of the wavepacket
we can observe different outcomes for the dynamics of the discrete state (see
Fig.~\ref{fig13}).

\section{Conclusions}
%

We have studied the properties of wave transmission in the
presence of a localized nonlinear defect described by the
generalized discrete Fano-Anderson model. This model provides a
simple generalization of the well-known problem of Fano resonances
to the nonlinear case, and it can be employed to describe the
so-called nonlinear Fano resonance effect. We have studied, both
analytically and numerically, the main features of the resonant
transmission for continuous waves and pulses, and we have shown
that the Fano resonance is observed as a specific feature in the
transmission coefficient as a function of frequency and/or
intensity. In particular, the presence of nonlinearity makes
the resonance more robust and it can broaden substantially the
parameter region where the resonant transmission is observed, by
adjusting either the wave frequency or the intensity to satisfy
the resonant conditions. Moreover, we have found the conditions
when the Fano resonance is associated with the bistable
transmission.

We would like to emphasize that many of the effects described in
this paper can find applications in a variety of different
nonlinear physical systems, including the transmission of
straight and curved waveguides
coupled to localized defect modes in photonic crystals. Moreover,
a number of interesting effects already observed in some of the
nonlinear systems can be identified as the manifestation of the
nonlinear Fano resonances. In particular, we believe that the
resent observation of excitonic optical bistability in
%
$\mathrm{Cu_2O}$~\cite{gdgbhhdghsmb04} can be explained in terms
of the nonlinear Fano resonance which could shed additional light on the
physics of these experimental observations.

\section*{Acknowledgements}

This work has been supported by the Australian Research Council.
We thank B. Malomed, S. Fan and M. Soljacic for useful discussions. A part
of this work has been conducted under the research projects of the
ARC Centre of Excellence CUDOS.


\begin{thebibliography}{11}

\bibitem{uf61} U. Fano, Phys. Rev. \textbf{124}, 1866 (1961).

\bibitem{junads94} J.U. N\"ockel and A.D. Stone, Phys. Rev. B \textbf{50}, 17415 (1994).

\bibitem{kkhaskyi02} K. Kobayashi, H. Aikawa, S. Katsumoto, and Y. Iye, Phys. Rev. Lett. \textbf{88}, 256806 (2002).

\bibitem{goeres00} J.~G\"ores, D.~Goldhaber-Gordon, S.~Heemeyer, M.~A.~Kastner, H.~Shtrikman, D.~Mahalu, and U.~Meirav,
Phys.~Rev.~B {\bf 62}, 2188 (2000).

\bibitem{bulka01} B.R.~Bulka and P.~Stefanski, Phys.~Rev.~Lett. {\bf 86}, 5128 (2001).

\bibitem{spinfilters} M.E. Torio, K. Hallberg, S. Flach, A.E. Miroshnichenko, and M. Titov, Eur. Phys. J. B {\bf 37}, 399 (2004);
A.A. Aligia and L.A. Salguero, Phys. Rev. B \textbf{70}, 075307
(2004).

\bibitem{shfjdj02} S. Fan and J.D. Joannopoulos, Phys. Rev. B {\bf 65}, 235112 (2002).

\bibitem{shf02} S.H. Fan, Appl. Phys. Lett. \textbf{80}, 908 (2002).

\bibitem{mfyshfms03} M.F. Yanik, S.H. Fan, and M. Soljacic, Appl. Phys. Lett. \textbf{83}, 2739 (2003).

\bibitem{arcjfy03} A.R. Cowan and J.F. Young, Phys. Rev. E \textbf{68}, 046606 (2003).

\bibitem{shfwsjdj03} S.H. Fan, W. Suh, and J.D. Joannopoulos, J. Opt. Soc. Am. B \textbf{20}, 569 (2003).

\bibitem{vljpv04} V. Lousse and J.P. Vigneron, Phys. Rev. B \textbf{69}, 155106 (2004).

\bibitem{etpfb93} E. Tekman and P.F. Bagwell, Phys. Rev. B \textbf{48}, 2553 (1993).

\bibitem{swksk01} S.W.~Kim and S.~Kim, Phys. Rev. B \textbf{63}, 212301 (2001).

\bibitem{sfaemvfmvf03} S. Flach, A.E. Miroshnichenko, V. Fleurov, and M.V. Fistul,
Phys. Rev. Lett. \textbf{90}, 084101 (2003).

\bibitem{mahan} G.D. Mahan, {\em Many-Particle Physics} (New York, Plenum Press, 1993).

\bibitem{aemsfbm03} A.E. Miroshnichenko, S. Flach, and B.A. Malomed, Chaos {\bf 13}, 874 (2003).

\bibitem{mingaleev} S.F. Mingaleev and Yu.S. Kivshar, J. Opt. Soc. Am. B {\bf 19}, 2241 (2002).

\bibitem{lipkin} H. J. Lipkin, {\em Quantum Mechanics: New Approaches to Selected Topics} 
(North-Holland, Amsterdam, 1973).

\bibitem{gibbs} H. M. Gibbs, {\em Optical Bistability: Controlling Light with Light}, (Academic Press, New-York, 1985).

\bibitem{bnzaas90} C. J. Joachain, {\em Quantum Collision Theory} (North-Holland, Amsterdam, 1975).

\bibitem{gdgbhhdghsmb04} G. Dasbach, G. Baldassarri H\"oger von  H\"ogersthal, D. Fr\"olich, H. Stolz, and M. Bayer, Phys. Rev. B \textbf{70}, 121202(R) (2004).


\end{thebibliography}
\end{document}